\documentclass[%
 reprint,
 amsmath,amssymb,
 aps,prl
]{revtex4-2}

\usepackage{graphicx}
\usepackage{dcolumn}

\usepackage[utf8]{inputenc}
\usepackage[T1]{fontenc}
\usepackage{mathptmx}
\usepackage{etoolbox}
\usepackage{bbm,bm,braket,amsmath,amssymb,amsfonts}

\def\one{\mathbbm{1}}

\begin{document}

\preprint{APS/123-QED}

\title{Reversible Cellular Automata as Integrable Interactions Round--a--Face:\\
Deterministic, Stochastic, and Quantized}

\author{Toma\v z Prosen}
\affiliation{%
Faculty of mathematics and physics, University of Ljubljana,
Jadranska 19, SI-1000 Ljubljana, Slovenia}

\date{\today}

\begin{abstract}
A family of reversible deterministic cellular automata, including the rules 54 and 201 of [Bobenko et al., Commun. Math. Phys. {\bf 158}, 127 (1993)] as well as their kinetically constrained quantum (unitary) or stochastic deformations, is shown to correspond to integrable Floquet circuit models with local interactions round-a-face.
Using inhomogeneous solutions of the star-triangle relation with a one or two dimensional spectral parameter, changing their functional form depending on the orientation, we provide an explicit construction of the transfer matrix and establish its conservation law and involutivity properties. Integrability is independently demonstrated by numerically exploring the spectral statistics via the Berry-Tabor conjecture. Curiously, we find that the deformed rule 54 model generically possesses no other local conserved quantities besides the net soliton current.
\end{abstract}

\maketitle


\emph{Introduction.--} 
Exactly solved models based on star-triangle (aka Yang-Baxter) equation represent the main cornerstone of statistical mechanics in two dimensions (2d) \cite{baxter,sierrabook}. 
These solutions provide nontrivial mappings from equilibrium physics in 2d to non-equilibrium (time-dependent) quantum and stochastic models in one dimension. This way one relates the Bethe ansatz for the spectra and eigenfunctions of quantum spin chains and stochastic simple exclusion processes (integrable interacting Markov chains) to integrable vertex models.

It is however very difficult and often even impossible to obtain explicit time-dependent results in typical integrable interacting models. Recently, a deterministic interacting particle model has been thoroughly studied~\cite{rule54review,prosen2016integrability,prosen2017exact,inoue2018two,gopalakrishnan2018facilitated,gopalakrishnan2018operator,gopalakrishnan2018hydrodynamics,klobas2019timedependent,alba2019operator,friedman2019integrable,klobas2020matrix,klobas2020exact,klobas2020space,klobas2021entanglement,klobas2021exactII}, namely the reversible cellular automaton (RCA) rule 54 of classification~\cite{bobenko1993two}. Particular striking and useful was the discovery of an 
exact time-dependent matrix-product ansatz for dynamics of observables which allows for the proof~\cite{klobas2019timedependent} of coexistence of ballistic (convective) and diffusive (conductive) transport, beyond hydrodynamic assumptions~\cite{GHDreview}.
The key feature which allows for further insight into dynamics as compared to generic integrable systems seems to be a constant scattering shift among quasiparticle excitations which all
propagate at the maximal bare velocity (cf. similar features observed recently in folded XXZ model \cite{Pozsgay2021,Lenart1,Lenart2} or q-boson model \cite{pozsgay2016realtime}).

However, despite a long standing effort it has not been established if and how the rule 54 RCA relates to Yang-Baxter integrability, whereas the connection has been understood a while ago \cite{inoueBBS} for a `less local' class of RCA, namely the box-ball systems \cite{BBS}. The locality and relativistic-look-a-like structure of the update rule of RCA 54 might suggest that one should look for the connection to integrability within the framework of interaction round-a-face (IRF) \cite{baxter,sierrabook,Pasquier,Sierra2,Pearce} rather than the more common vertex models. 

This is exactly what we attempt and achieve in this Letter
by embedding (deforming) RCA rules 54 and 201 into a large four-parametric families of 
kinetically constrained quantum (Floquet circuit) or classical stochastic (Markov circuit) models. We find novel inhomogeneous (or anisotropic) solutions of the star-triangle equation where the $R$-matrices take functionally different forms depending on the orientation of the diagram. We build the transfer matrix generating an extensive set of conserved and mutually commuting operators, which seem to be, however, manifestly nonlocal for the deformed rule 54.
This result provides a basis to study broad classes of integrable kinetically constrained models.

\emph{Local IRF circuits.--}
Consider a Hilbert space of states $\mathcal H = (\mathbb C^2)^{\otimes 2N} = \mathbb C^{2^{2N}}$ of a chain of $2N$ qubits and define a class of linear
dynamical systems over $\mathcal H$
with the generator
\begin{equation}
\mathcal U = \mathcal U^{\rm o}\mathcal U^{\rm e}\,, \quad
\mathcal U^{\rm o} = \prod_{x=1}^N F_{2x-1}\,, \quad
\mathcal U^{\rm e} = \prod_{x=1}^N F_{2x}\,,
 \label{IRFC}
\end{equation}
where
$F_x$ denotes a 3-site map embedded into ${\rm End}(\mathcal H)$ acting nontrivially on site $x$ with the control on sites $x-1$ and $x+1$
\begin{equation}
F_x = \sum_{i,j,k,l=0}^1 (f_{kl})_{i}^{j}\, 
\ket{k}\!\bra{k}_{x-1}\, \ket{i}\!\bra{j}_{x}\, 
\ket{l}\!\bra{l}_{x+1}\,.
\label{3gate}
\end{equation}
$F_x$ is completely specified in terms of $4$, $2\times 2$ matrices $\{f_{kl}\in {\rm End}(\mathbb C^2)\}_{k,l=0,1}$.
The propagator 
\begin{equation}
\left(\mathcal U^t\right)
_{i_1,\,i_2\,\ldots\,i_{2N}}^{j_1,j_2\ldots j_{2N}}
= \vcenter{\hbox{\includegraphics[scale=0.4]{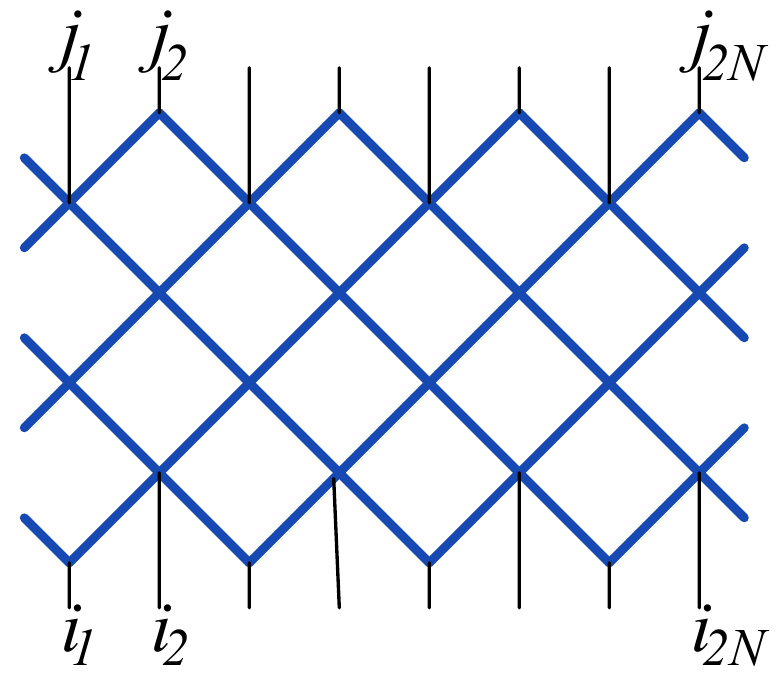}}},
\end{equation}
corresponds to a partition-sum of interactions round-a-face (IRF) model on a $2N\times 2t$ cylinder with face weights (gate) 
\begin{equation}
    (f_{kl})_{i}^{j} = \vcenter{\hbox{\includegraphics[scale=0.4]{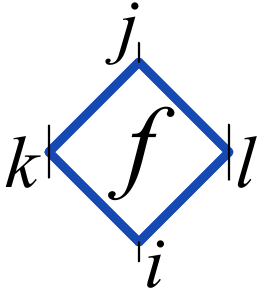}}}. 
    \label{face}
\end{equation}    
Here and below we designate rows/columns of matrices with sub/superscript binary-digit 
indices\footnote{$A_{i_1,i_2\ldots}^{j_1,j_2\ldots}\equiv \bra{i_1,i_2\ldots}A\ket{j_1,j_2,\ldots}$.}.
Periodic boundaries $x+2N\equiv x$
are assumed throughout.

\begin{figure}[t]
\includegraphics[scale=0.68]{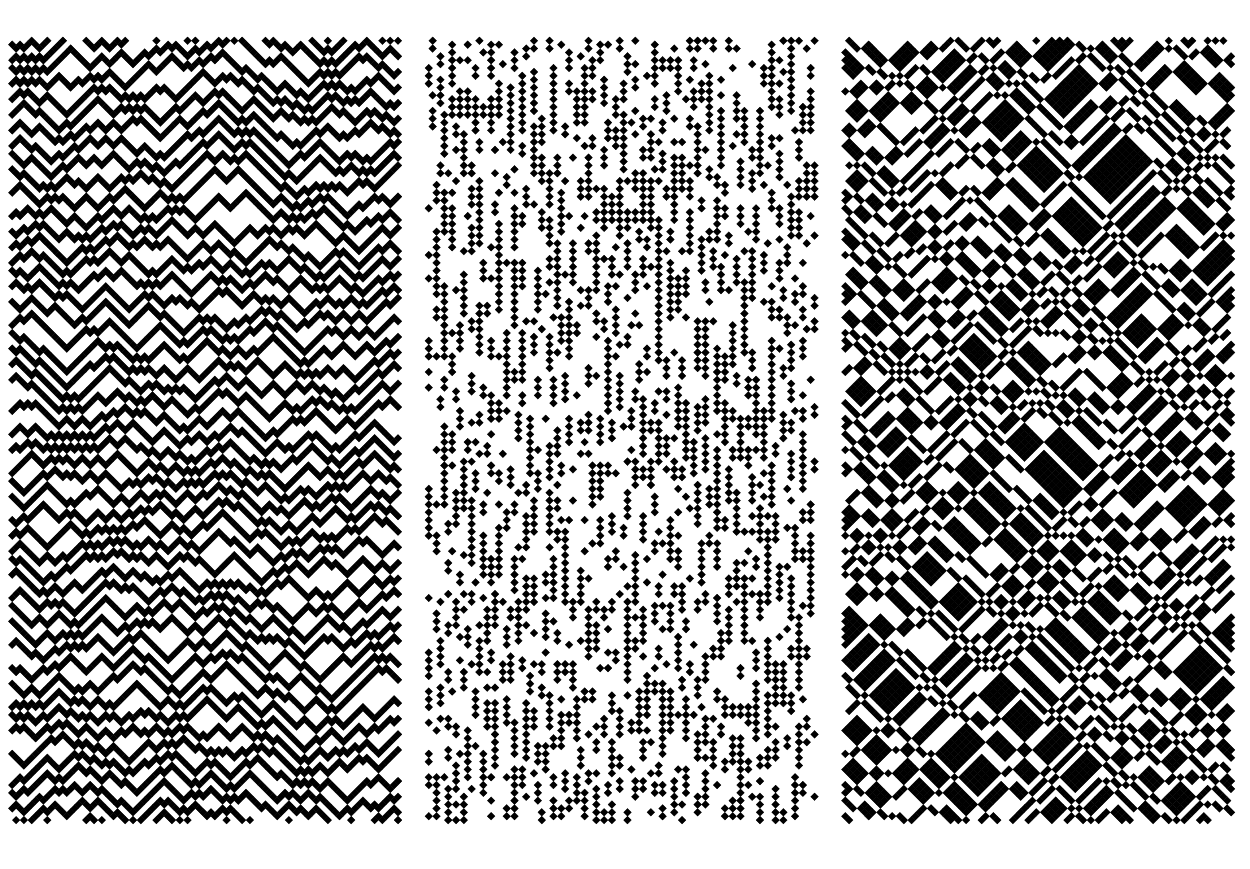}
\caption{\label{fig:SCA} 
Exemplar stationary state trajectories of SCAs: stochastic deformations of rule 54 (left), 201 (middle), 150 (right), all for $\alpha=0.8$, $\delta=0.65$ and lattice halfsizes $(N,t)=(50,100)$ (time runs downwards). While the first two models are Yang-Baxter integrable, the third seems not.}
\end{figure}

We immediately find three distinct interesting physical contexts of IRF dynamics (\ref{IRFC},\ref{3gate}): (i) When $f_{kl}\in\{\one,X\}$ (where $X$ is a Pauli/NOT gate) the circuit (\ref{IRFC}) corresponds to
\emph{deterministic reversible cellular automaton} (RCA), one of $2^4=16$ time-reversal symmetric cases classified in Ref.\cite{bobenko1993two}. Specifically, $\mathcal U$ maps basis states $\ket{i_1,\ldots,i_{2N}}$, $i_x\in\{0,1\}$, to basis states, so dynamics can be fully characterised in terms of trajectories of configurations 
$\ket{i^t_1,\ldots,i^t_{2N}} = \mathcal U^t 
\ket{i^0_1,\ldots,i^0_{2N}}$.
Notable examples considered here are:
\begin{eqnarray}
&\textrm{rule 54}: 
\quad &
f_{00}=\one, \quad f_{01}=f_{10}=f_{11}=X,\label{54}\\
&\textrm{rule 201}:
\quad &
f_{00}=X, \quad f_{01}=f_{10}=f_{11}=\one,\label{201}\\
&\textrm{rule 150}:
\quad &
f_{00}=\one, \quad f_{01}=f_{10}=X,\;\;f_{11}=\one.\label{150}
\end{eqnarray}
While the rule 54 is a paradigmatic solvable example of interacting deterministic dynamics with coexisting ballistic and diffusive transport \cite{klobas2019timedependent,rule54review}, rule 201 represents a classical deterministic~\cite{wilkinson2020exact} and Floquet~\cite{FloquetPXP1,FloquetPXP2}
variant of PXP model~\cite{PXP,Lukin} of kinetically constrained dynamics~\cite{garrahan2018aspects}, and rule 150
is a classical deterministic version of the XOR-Fredrickson-Andersen model~\cite{Causer2020} which can be interpreted in terms of non-interacting dynamics~\cite{wilkinson2021}.
(ii) When $f_{kl}\in {\rm U}(2)$ are \emph{unitary}, also the many body map (\ref{IRFC}) is unitary and corresponds to a unitary quantum local Floquet circuit or \emph{quantum cellular automaton} (QCA)~\cite{Farrelly}, see e.g.~\cite{Carr,TP21} for recent discussion of QCA of this type.
(iii) When $f_{kl}$ are \emph{stochastic} matrices, 
$(f_{kl})_i^j \in \mathbb R_{\ge 0}$, $\sum_i (f_{kl})_i^j = 1$, the 
map
(\ref{IRFC}) is also stochastic and corresponds to a local Markov chain circuit or \emph{stochastic cellular automaton} (SCA), see Fig.~\ref{fig:SCA} for typical sections of long-time trajectories.

In this Letter we shall consider kinetically constrained deformation of RCA (\ref{54}-\ref{150}) where $f_{00}$ is replaced by
\begin{equation}
    f_{00} = 
    \begin{pmatrix}
    \alpha & \beta \\
    \gamma & \delta
    \end{pmatrix}\,, \label{u00}
\end{equation}
and $\alpha,\beta,\gamma,\delta\in\mathbb C$ are arbitrary mixing parameters. In other words, the only non-deterministic local process in the map (\ref{3gate}) happens when neigbouring sites are both in state $0$: $\ket{000} \to \alpha \ket{000} + \beta\ket{010}$, $\ket{010} \to \gamma \ket{000} + \delta\ket{010}$.
We stress that all algebraic integrable structures discovered here apply for arbitrary matrix (\ref{u00}), being either unitary, stochastic, or even general complex.

\begin{figure}[t]
\includegraphics[scale=0.64]{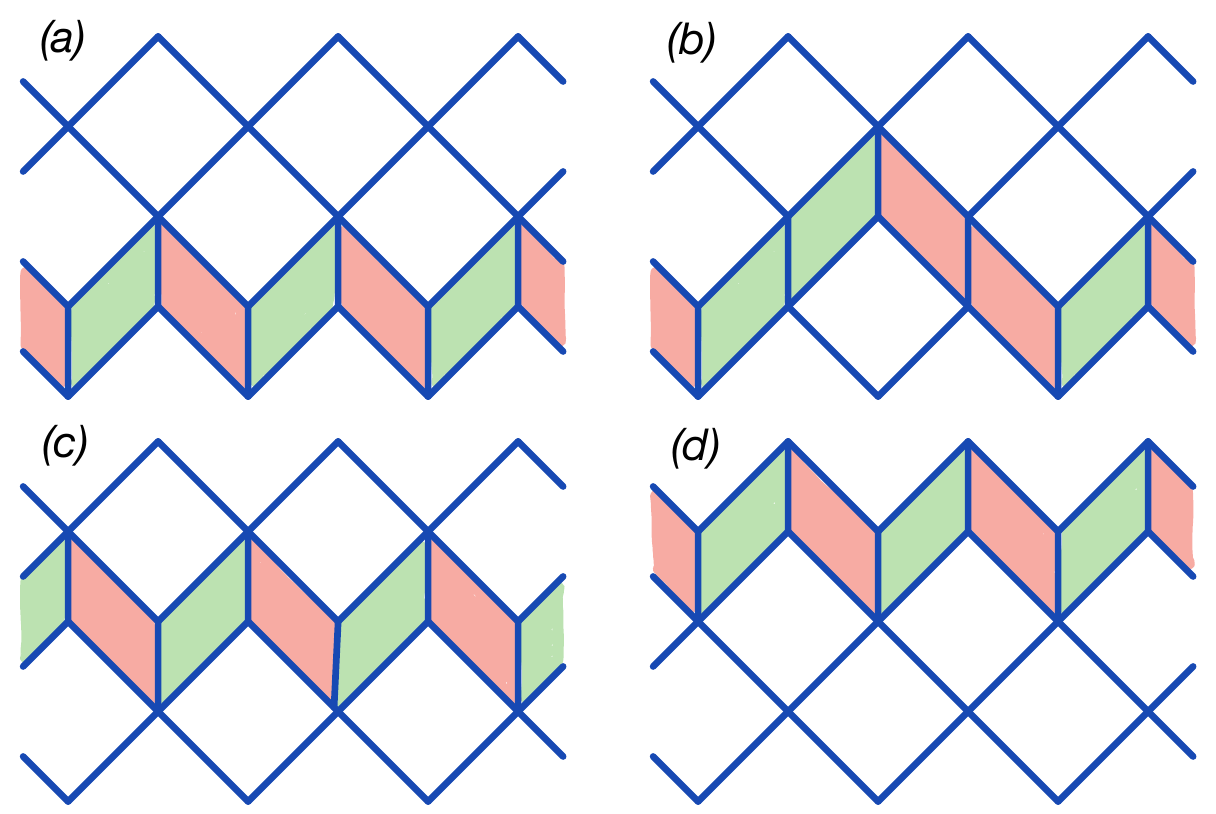}
\caption{\label{fig:ConsTM} 
Conservation law property (\ref{cl}) of the transfer matrix (\ref{TM}) via implication series of identical circuits $(a)\Rightarrow(b)\Rightarrow(c)\Rightarrow(d)$.
}
\end{figure}

\emph{Integrability of deformed rule 54.--}
We first consider the deformed rule 54 dynamics with (\ref{54},\ref{u00}). Taking $4\times 4$ matrices parametrizing the face weights 
$L_{i_1,i_2}^{j_1,j_2} = \vcenter{\hbox{\includegraphics[scale=0.3]{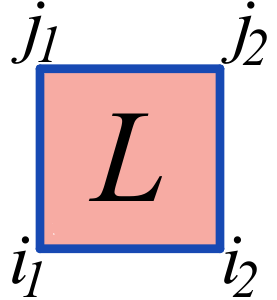}}}$,
$M_{i_1,i_2}^{j_1,j_2} = \vcenter{\hbox{\includegraphics[scale=0.3]{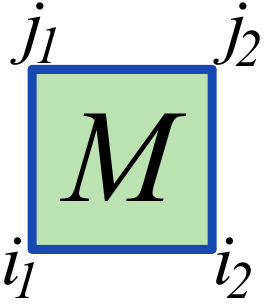}}}$, we
require them to satisfy the star-triangle (aka checkerboard Yang-Baxter) equation where the propagator (\ref{face}) plays the role of the $R-$matrix:
\begin{eqnarray}
\vcenter{\hbox{\includegraphics[scale=0.3]{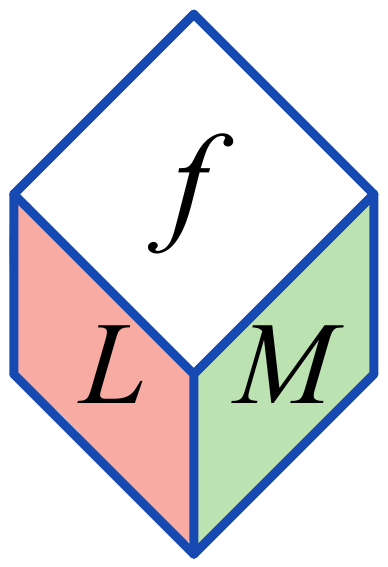}}} &=& \vcenter{\hbox{\includegraphics[scale=0.3]{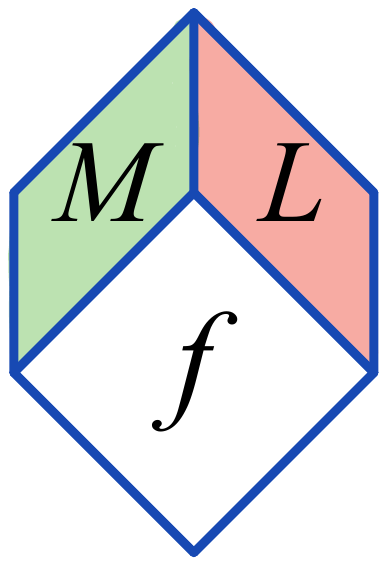}}} \label{st}\\
\sum_s L_{k,l}^{j,s}\,(f_{jn})_s^i\,M_{l,m}^{s,n} &=&
\sum_s M_{k,s}^{j,i}\,(f_{km})_l^s\,L_{s,m}^{i,n}\,. \nonumber
\end{eqnarray}
Remarkably, this system of $2^6$ quadratic equations admits a non-trivial solution for the 32 components of Lax matrices $L,M$ (using the Gr\" obner basis algorithm of Mathematica):
\begin{eqnarray}
L(\lambda) &=&
\left(
\begin{array}{cccc}
 1 & 0 & 1 & 0 \\
 1 & 0 & 1 & 0 \\
 0 & \lambda^2 & 0 & \lambda  \\
 0 & \lambda  & 0 & 1 \\
\end{array}
\right)\,, \label{Lsol}\\
M(\lambda) &=&
\left(
\begin{array}{cccc}
 \alpha  & \beta  & 0 & 1 \\
 0 & 1/\lambda & 0 & 1/\lambda \\
 \gamma  & \delta  & 1/\lambda & 0 \\
 1 & 0 & 1/\lambda & 0 \\
\end{array}
\right)\,.\quad\label{Msol}
\end{eqnarray}
Exploring the gauge invariance of the solution of (\ref{st}) under
\begin{equation}
L_{i_1,i_2}^{j_1,j_2} \to c \frac{a_{i_1} b_{j_1}}{a_{i_2} b_{j_2}} L_{i_1,i_2}^{j_1,j_2}\,,
\quad
M_{i_1,i_2}^{j_1,j_2}
\to \frac{1}{c}\frac{a_{i_1} b_{j_1}}{a_{i_2} b_{j_2}} M_{i_1,i_2}^{j_1,j_2}\,,
\label{gauge}
\end{equation} 
for arbitrary $a_i,b_i,c\in \mathbb C$, we chose representation  (\ref{Lsol},\ref{Msol}) that depends on a single spectral parameter $\lambda$ that cannot be gauged out. The solution family (\ref{Lsol},\ref{Msol}) is unique up to time reflection ($t\to-t$), $f_{00} \to f_{00}^{-1}$ (if invertible), or to space reflection ($x\to 2N+1-x$), 
$L_{i_1,i_2}^{j_1,j_2}\to M_{i_2,i_1}^{j_2,j_1}$,
$M_{i_1,i_2}^{j_1,j_2}\to L_{i_2,i_1}^{j_2,j_1}$. 
Clearly, there exists \emph{no} gauge in which $L$ and $M$ differ only by a value of the spectral parameter, so they are functionally different.
Componentwise defining a transfer matrix (TM) 
\begin{equation}
    \mathcal T_{\,i_1,\,i_2\ldots\,i_{2N}}^{j_1,j_2\ldots j_{2N}}
    (\lambda) =
    \prod_{x=1}^N 
    M_{i_{2x-1},i_{2x}}^{j_{2x-1},j_{2x}}(\lambda) L_{i_{2x},i_{2x+1}}^{j_{2x},j_{2x+1}}(\lambda)\,,
    \label{TM}
\end{equation}
as  $\mathcal T(\lambda) \in {\rm End}(\mathcal H)$,
manifestly invariant under gauge transformation (\ref{gauge}),
the star-triangle equation (\ref{st}) implies its conservation law property (see Fig.~\ref{fig:ConsTM} for diagrammatic proof)
\begin{equation}
\mathcal T (\lambda) \mathcal U =
\mathcal U \mathcal T (\lambda) \,.\label{cl}
\end{equation}

Further, we establish the involution property of the TM:
\begin{equation}
    [\mathcal T(\lambda),
    \mathcal T(\lambda')]=0,\quad\forall\lambda,\lambda' \in\mathbb C\setminus\{0\}\,. \label{inv}
\end{equation}
Suppose there exist a dual (horizontal) 3-site control gate
\begin{equation}
    (g_{kl})_i^j \equiv (\tilde{g}_{ij})_k^l =  \vcenter{\hbox{\includegraphics[scale=0.4]{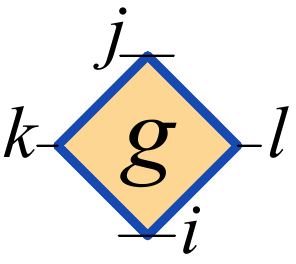}}}. \label{gace}
\end{equation}
satisfying the sideways star-triangle equations 
\begin{eqnarray}
 \vcenter{\hbox{\includegraphics[scale=0.3]{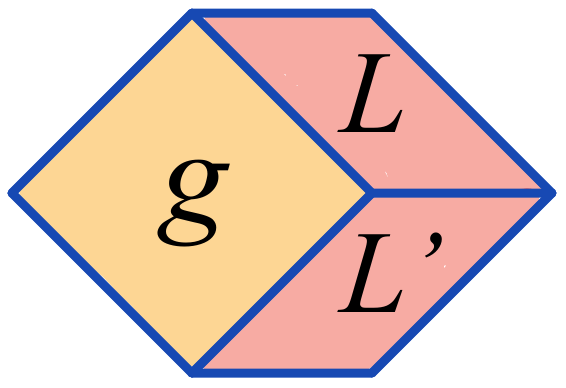}}} &=& \vcenter{\hbox{\includegraphics[scale=0.3]{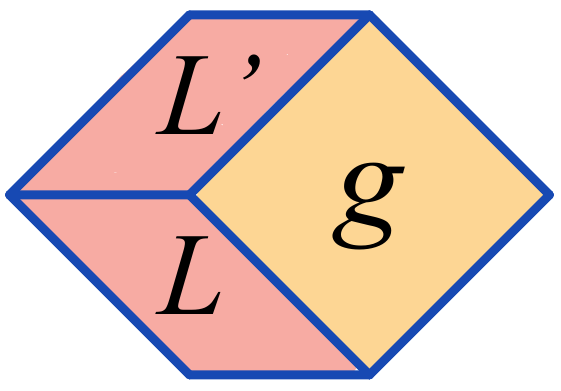}}} \label{gLL}\\
 \sum_s 
 (g_{js})_k^i\,L_{s,m}^{i,n}(\lambda)\,L_{k,l}^{s,m}(\lambda') &=&
\sum_s L_{j,s}^{i,n}(\lambda')\,L_{k,l}^{j,s}(\lambda)\,(g_{sm})_l^n\,,
\nonumber
\end{eqnarray}
for $L$-matrices, and similarly intertwining the $M$-matrices:
\begin{eqnarray}
\vcenter{\hbox{\includegraphics[scale=0.3]{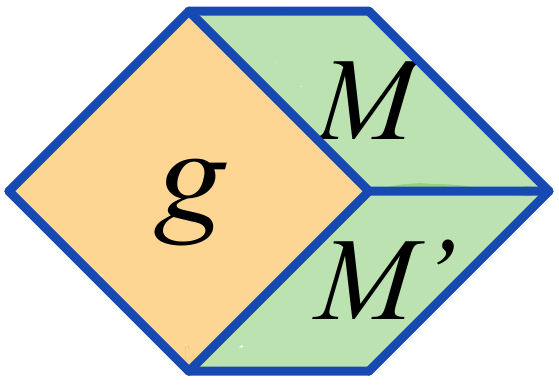}}} &=& \vcenter{\hbox{\includegraphics[scale=0.3]{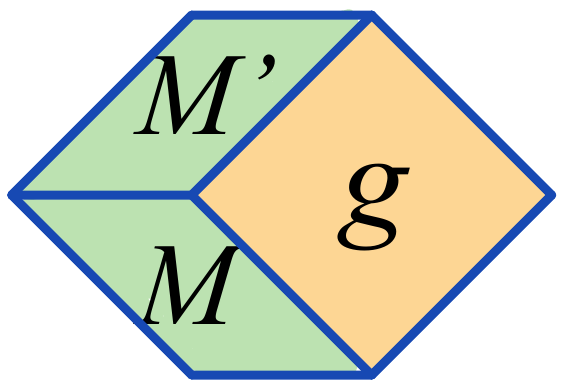}}} \label{gMM} \\
\sum_s (g_{js})_k^i\,M_{s,m}^{i,n}(\lambda)\,M_{k,l}^{s,m}(\lambda') &=&
\sum_s
M_{j,s}^{i,n}(\lambda')\,M_{k,l}^{j,s}(\lambda)\,(g_{sm})_l^n\,,
\nonumber
\end{eqnarray}
where  $L\equiv L(\lambda)$,
$L'\equiv L(\lambda')$, 
$M\equiv M(\lambda)$
$M'\equiv M(\lambda')$.
Then, assuming invertibility of all $2\times 2$ matrices $\tilde{g}_{ij}$ (\ref{gace}) one writes a simple diagrammatic prof of involutivity (\ref{inv}):
\begin{eqnarray}
&& \vcenter{\hbox{\includegraphics[scale=0.36]{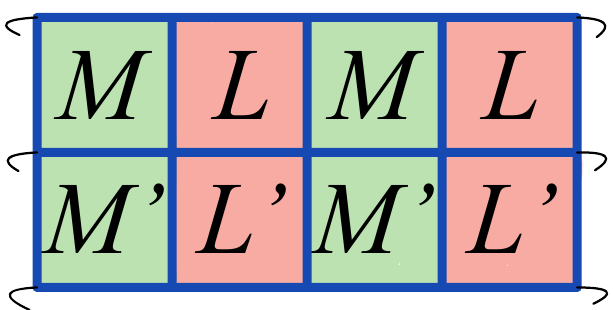}}} =
\vcenter{\hbox{\includegraphics[scale=0.36]{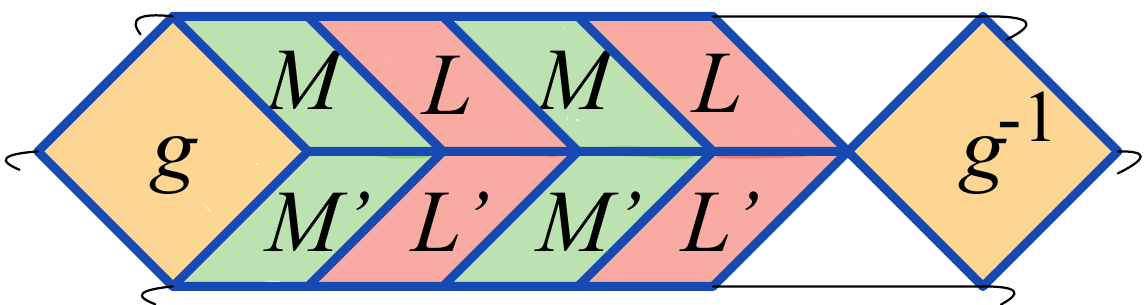}}}= \nonumber\\
&& = \vcenter{\hbox{\includegraphics[scale=0.36]{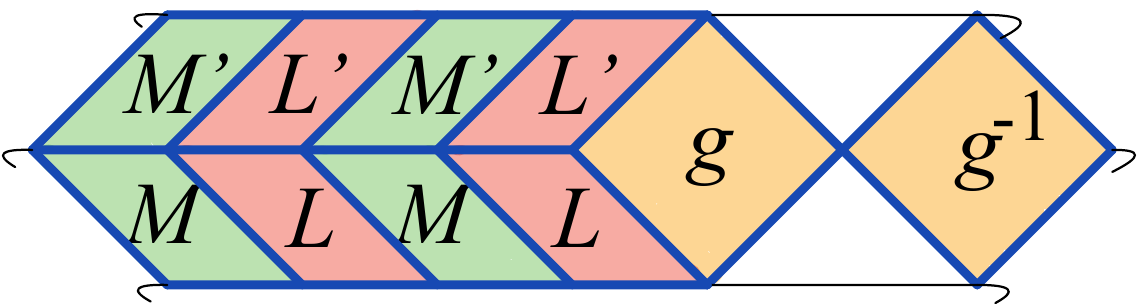}}} =
\vcenter{\hbox{\includegraphics[scale=0.36]{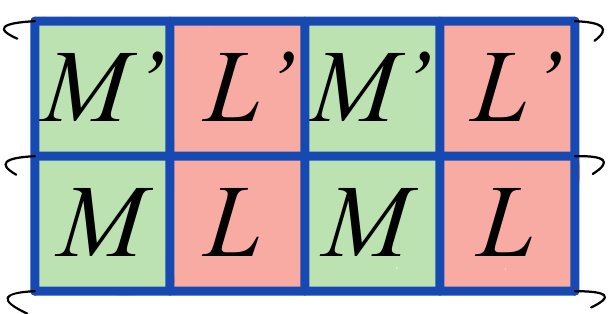}}}
\end{eqnarray}
where the inverted dual gate ``$g^{-1}$'' is built as (\ref{gace}) with 
$\tilde{g}_{ij}$ replaced by $\tilde{g}_{ij}^{-1}$. 
In fact, (\ref{gLL},\ref{gMM}) is a system of linear equations for the unknown \emph{spatial} gate $g$, having a solution (unique up to normalization)
\begin{equation}
    (g_{kl})_i^j = L_{k,j}^{i,l}(\lambda'/\lambda)\,.
\end{equation}
We note that, analogously, the \emph{temporal} IRF gate (\ref{face}) can be expressed in terms of the other Lax matrix
\begin{equation}
    (f_{kl})_i^j = M_{i,l}^{k,j}(\lambda=1)\,.
\end{equation}
The coefficients of the Laurent polynomial $\mathcal T(\lambda)$
provide an extensive set of, however non-local, conserved operators. We stress that $\mathcal T(\lambda)$ is nontrivial also in the deterministic case $\alpha=\delta=1$, $\beta=\gamma=0$ and hence gives the long sought Yang-Baxter transfer matrix of the rule 54 model. 

\emph{Integrability of deformed rule 201.--} Following exactly the same procedure for deformed rule 201 (\ref{201},\ref{u00}) we obtain, up to a gauge (\ref{gauge}) and spatial/temporal reflection, 
unique~\footnote{Provided we make some nondegeneracy assumption, otherwise we get two other solution families 
with additional zero weights.} Lax operators depending on a 2-component spectral parameter,
in terms of which we again express the spatial and temporal 
IRF gates:
\begin{eqnarray}
L(\boldsymbol{\lambda}) &=& M(\boldsymbol{\lambda})
\vert^{\alpha=\delta=1}_{\beta=\gamma=0}\,, \label{L201}\\
M(\boldsymbol{\lambda})&=&
\left(
\begin{array}{cccc}
 \alpha & \beta/\lambda_1 & \lambda_1 & 0 \\
 \lambda_1 & 0 & \lambda_1^2 & 0 \\
 \gamma/\lambda_1 & \delta/\lambda_1^2 & 0 & 1 \\
 0 & 1 & 0 & \lambda_2 \\
\end{array}
\right)\label{Lax201}\,, \label{M201}\\
 (f_{kl})_i^j &=& M_{i,l}^{k,j}(1,1)\,, \\
 (g_{kl})_i^j &=&
 L_{k,j}^{i,l}(\lambda'_1/\lambda_1,\lambda'_2/\lambda_2)\,.
\end{eqnarray}
These objects in turn define a 2-parametric conserved (\ref{cl}) TM $\mathcal T(\boldsymbol{\lambda})$, 
cf.~(\ref{TM}), satisfying involutivity (\ref{inv}).

Attempting to repeat the procedure for deformed rule 150 (\ref{150},\ref{u00}) we failed to find nontrivial solutions to star-triangle relation (\ref{st}). Thus we conclude that, while rules 54 and 201 can be considered embedded in large multiparametric integrable families, the noninteracting rule 150 becomes instantly non-integrable when deformed.

Alternatively, it may be convenient to use a notation closer to Baxter~\cite{baxter} and parametrize $L$, $M$ in terms of quads of $2\times 2$ matrices $\ell_{kl},m_{kl}$, via 
$(\ell_{kl})_i^j \equiv L_{i,l}^{k,j}$, 
$(m_{kl})_i^j \equiv M_{i,l}^{k,j}$ and defining embedded 3-site local operators $L_x$ or $M_x$ via (\ref{3gate}) on extended state space $\mathcal H'=(\mathbb C^2)^{\otimes(2N+2)}$ (including a pair of sites $x=0,2N+1$),
with $f_{kl}$ replaced by $\ell_{kl}$ or $m_{kl}$, respectively. Then, the star-triangle Eq.~(\ref{st}) reads $F_{x+1} M_{x} L_{x+1} = L_x M_{x+1} F_x$ (and analogously for (\ref{gLL},\ref{gMM})).
TM (\ref{TM}) now reads as an ordered operator product equipped with projectors
$P_{x,y}=\sum_i \ket{i}\!\bra{i}_x \ket{i}\!\bra{i}_y$
and tracing out the two auxiliary sites,
\begin{equation}
    \mathcal T = {\rm tr}_{0,2N+1}(P_{1,2N+1} M_1 L_2 M_3 L_4 \cdots M_{2N-1} L_{2N} P_{0,2N})\eta,
    \label{TM2}
\end{equation}
where $\eta$ is a lattice shift operator over $\mathcal H$ defined as 
$\eta\ket{i_1 i_2\ldots i_{2N}} = \ket{i_2 \dots i_{2N}i_1}$. In case of unitary deformations (\ref{u00}) the matrices $\ell_{kl}(\boldsymbol{\lambda}),m_{kl}(\boldsymbol{\lambda})$, and hence $L_x,M_x$, are unitary exactly when the spectral parameters(s) is(are) unimodular $|\lambda_i|=1$. Due to boundary terms in (\ref{TM2}) one would perhaps expect finite size breaking of unitarity of TM. Nevertheless, we find by explicit computation that TM is exactly unitary
$\mathcal T(\boldsymbol{\lambda})^\dagger \mathcal T(\boldsymbol{\lambda})|_{|\lambda_i|=1}=\one$, for both cases (\ref{Lsol},\ref{Msol}), and
(\ref{L201},\ref{M201}).

\emph{Local conservation laws and level statistics.--} In order to study physics of (deformed) RCA it is natural to investigate the structure of (quasi)local conserved charges or currents \cite{ilievskireview,pozsgayPRL2020,pozsgayPRX2020}. We note first that all
RCA (\ref{54},\ref{201},\ref{150}) with general deformation (\ref{u00}), 
preserve a diagonal ${\rm U}(1)$ charge which can be interpreted as a net soliton current (the density of right movers minus the density of left movers), $Z$ is a diagonal Pauli matrix:
\begin{equation}
    \mathcal J = 
    \sum_{x=1}^N 
    (Z_{2x-1}Z_{2x} - Z_{2x}Z_{2x+1}),\quad
    [\mathcal J,\mathcal U]=0\,.
    \label{J}
\end{equation}
For stochastic deformations this can also be observed inspecting Fig.~\ref{fig:SCA}.
As TM (\ref{TM2}), for generic unitary deformation, does not reduce to a shift at any specific value of the spectral parameter, its log-derivatives
$\partial_{\boldsymbol{\lambda}}\log \mathcal T(\boldsymbol{\lambda})=[\mathcal T(\boldsymbol{\lambda})]^\dagger \partial_{\boldsymbol{\lambda}} \mathcal T (\boldsymbol{\lambda})$ will not yield local operators for any unimodular $\boldsymbol{\lambda}$.
Yet, we may expect that the latter are \emph{quasilocal} charges, in full analogy to established quasilocal charges related to six-vertex model~\cite{ilievskireview}.
We have furthermore checked empirically for the existence of any strictly local conserved charges by adapting the Algorithm 2, Sec.~7 of Ref.~\cite{TP07} from the $2$-qubit operator gate to $3$-qubit control-operator gate
$\hat{W}_x(a) = F^{-1}_x a F_x$. Remarkably, we have indeed found \emph{no} additional local conserved charge for support size up to $r=5$ consecutive sites, for the deformed rule 54. For deformed rule 201, we found one charge
\emph{even} under $x\to 2N+1-x$ (in addition to an \emph{odd} one (\ref{J})) for $r=3$, and additional two even and two odd charges for $r=5$. Notably, all the \emph{local} charges $\mathcal Q$ for deformed rule $201$ are diagonal (can be written solely in terms of $Z_x$) and are unrelated (Hilbert-Schmidt orthogonal) to TM ${\rm tr}\mathcal Q \mathcal T\!(\boldsymbol{\lambda}) \equiv 0$.
Unsurprisingly, we have found no additional local charges for the nonintegrable deformed rule 150. 

\begin{figure}[t]
\begin{center}

\includegraphics[scale=0.27]{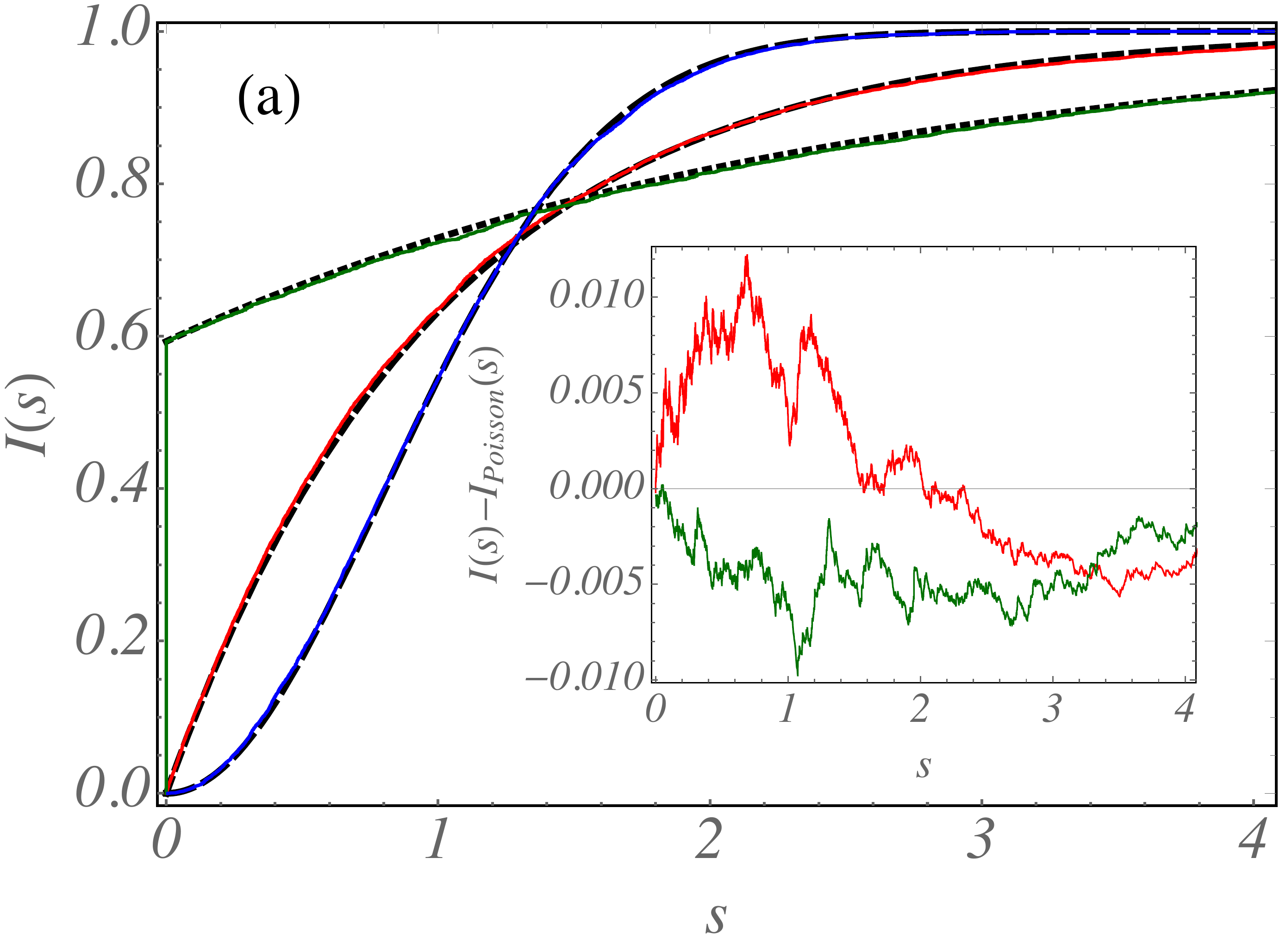}

\hspace{5mm}\includegraphics[scale=0.37]{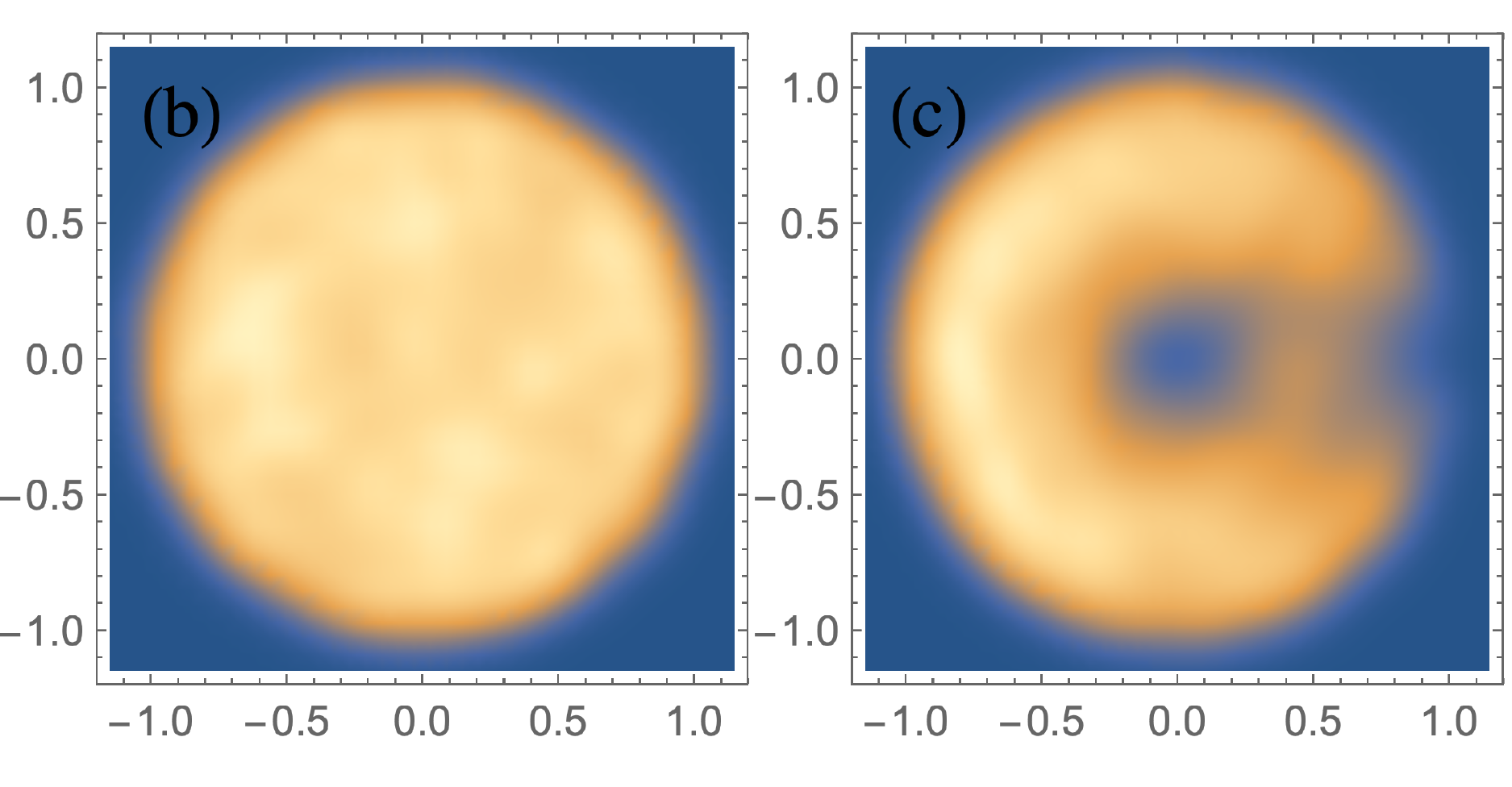}

\end{center}
\vspace{-4mm}

\caption{\label{fig:Is} 
Spectral statistics:
(a) Cumulative level spacing distributions $I(s)$ for quantum deformed rule 54 (red), 201 (green), 150 (blue), with unitary deformation
$\alpha=e^{{\rm i}(\phi+\nu)}\cos\omega$,
$\beta=e^{{\rm i}(\phi+\mu)}\sin\omega$,
$\gamma=-e^{{\rm i}(\phi-\mu)}\sin\omega$,
$\delta=e^{{\rm i}(\phi-\nu)}\cos\omega$, 
$\omega=0.5$, $\phi=1.3$, $\nu=0.7$, $\mu=0.6$.
Poissonian $I_{\rm Poisson}=1-e^{-s}$ 
(Wigner-Dyson $I_{\rm WD}(s) = 
1-e^{-\pi s^2/4}$) distribution is plotted with short (long) dashed black curve. The case of rule 201 has a macroscopic degeneracy of $I_0 = I(s\to 0) = 0.593$, while the dotted black curve designates an effective Poissonian $I_0 + (1-I_0)e^{-(1-I_0)s}$.
The inset shows fine (below $1\%$) deviations from the Poisson model.
(b,c) Probability density of complex spacing ratios for deformed rule 54 (b), and 150 (c) with complex
(non-unitary\&non-stochastic) deformation:
$\alpha=1+ 0.5{\rm i}$, $\beta=-0.5+0.5{\rm i}$, $\gamma=0.5-0.3{\rm i}$, $\delta=0.6+0.7{\rm i}$.
The average radial and angular spacing ratios (see \cite{LucasPRX} for definitions) are:
(b) $\langle r\rangle = 0.669$, $\langle\cos\theta\rangle=0.002$ (Poissonian values $2/3, 0$),
(c) $\langle r\rangle = 0.719$, $\langle\cos\theta\rangle=-0.177$ (${\rm AI}^\dagger$ random matrix ensemble estimates $0.722$, $-0.185$).
All spectra contain 10800 levels each ($N=9$, sector with fixed quasi-momentum $K=1$ and soliton current $J=0$).
}
\end{figure}

It seems rather striking to suggest an existence of a completely integrable extended locally interacting model with a single local conserved quantity. One may wonder if such a model still obeys the standard conjectures, say if eigenvalues of Floquet propagator $\mathcal U$ represent an uncorrelated Poisson process (Berry-Tabor conjecture \cite{BerryTabor,Marklof}). Given $N$, we have diagonalized the matrix of $[\mathcal U]_K$ projected to fixed quasi-momentum $K\in\{0,\ldots,N-1\}$ sector of Hilbert space $\{\ket{\psi};\eta^2\ket{\psi}=e^{2\pi{\rm i}K/N}\ket{\psi}\}$. As $[\mathcal U]_K=\mathcal V^2$ can be written as a square~\cite{LucasPRB}, where $\mathcal V = [\eta \mathcal U^{\rm e}]_K$, we have thus analyzed the spectrum of $\mathcal V$ fixing as well the eigenvalue $J$ of the conserved soliton current $\mathcal J$. 
The numerical results are summarized in Fig.~\ref{fig:Is} showing the cumulative level spacing distribution $I(s)$, giving the probability that a random spacing between adjacent eigenphases of $\mathcal V$ (normalized to mean unit spacing) is smaller than $s$, for unitary (quantum) deformations of automata (\ref{54},\ref{150},\ref{201}). While generic Poisson spectrum was observed for deformed rule 54, a macroscopic degeneracy has been observed for deformed rule 201 (suggesting hidden non-abelian quantum symmetry of the model) again conforming to the Poisson model for non-degenerate levels. Finally, deformed rule 150 exhibits random matrix spectral statistics, suggesting it to be a quantum chaotic model. Fully compatible results have been found for spectral statistics of generic complex deformations of the RCAs in terms of analyzing complex spacing ratios~\cite{LucasPRX} which thus model integrable and chaotic non-Hermitian~\cite{Ueda} kinetically constrained systems.

\emph{Conclusions.--} This Letter provides a backbone of integrability for kinetically constrained interacting lattice dynamics, being either deterministic, stochastic or quantum. This immediately opens several avenues for further quest of more explicit results related to nonequilibrium dynamics of these models, e.g. identifying complete sets of quasiparticle excitations and fundamental scattering relations among them leading to the generalized hydrodynamic theory of transport \cite{GHDreview}, or targeting exact spatio-temporal correlation functions generalising the results~\cite{klobas2019timedependent,klobas2020exact,klobas2021entanglement} for the deterministic rule 54.

Finally, our model should also be of interest as a nontrivial exactly solved model in 2d statistical mechanics, either thinking of a homogeneous partition function composed of $f$- or $g$- IRF plaquettes, or a staggered partition function composed of columns of $L$- and $M$- plaquettes. Apparently, previous attempts to construct exactly solvable staggered IRF models could not go beyond free fermion models~\cite{Maillard}.

\begin{acknowledgments}
Illuminating discussions with F.~C.~Alcaraz, B.~Pozsgay and G.~Sierra are gratefully acknowledged. 
I thank H.~Katsura, V.~Pasquier and L.~Piroli for helpful remarks on the manuscript and to B.~Bu\v ca, J.~P.~Garrahan, K.~Klobas, M.~Medenjak, M.~Vanicat, and J.~W.~P.~Wilkinson, for collaboration on related projects. This work has been supported by the European Research Council under the
Advanced Grant No.\ 694544 -- OMNES, and by the Slovenian Research Agency
under the Program P1-0402.
\end{acknowledgments}



\bibliography{bibliography}
\end{document}